**Investigation of Growth-Induced Strain in Monolayer MoS$_2$ Grown by Chemical Vapor Deposition**

Siwei Luo[1,2], Conor P. Cullen[3], Gencai Guo[2], Jianxin Zhong[2]\*, Georg S. Duesberg[1,3]\*

[1]*Institute of Physics, EIT 2, Faculty of Electrical Engineering and Information Technology, Universität der Bundeswehr München, Werner-Heisenberg-Weg 39, 85577 Neubiberg, Germany*

[2] *Hunan Key Laboratory for Micro-Nano Energy Materials and Devices, Laboratory for Quantum Engineering and Micro-Nano Energy Technology, and School of Physics and Optoelectronics, Xiangtan University, Hunan 411105, People's Republic of China*

[3] *School of Chemistry, Centre for Research on Adaptive Nanostructures and Nanodevices (CRANN) and Advanced Materials and Bioengineering Research (AMBER), Trinity College Dublin, Dublin 2, Ireland*

\*Corresponding author.
E-mail address:

georg.duesberg@unibw.de (Georg S. Duesberg)

jxzhong@xtu.edu.cn. (Jianxin Zhong)

**Abstract**

Two-dimensional materials such as transitional metal dichalcogenides exhibit unique optical and electrical properties. Here we report on the varying optical properties of CVD grown $MoS_2$ monolayer flakes with different shapes. In particular, it is observed that the perimeter and the central region of the flakes have non-uniform photoluminescence (PL) energy and intensity. We quantified these effects systematically and propose that thermally induced strain during growth is the origin. The strain relaxation after transfer of the $MoS_2$ flakes supports this explanation.

Detailed investigations of the spatial distribution of the PL energy reveal that depending on the shape of the $MoS_2$ flakes, the width of the strain field is different. Thus, our results help to elucidate the fundamental mechanisms responsible for the differences in PL and Raman signals between the perimeter region and the center region of monolayer $MoS_2$ and suggest that the induced strain plays an important role in the growth of monolayer materials.

## I. INTRODUCTION

Two-dimensional (2D) materials, including graphene, black phosphorene (BP) and transition metal dichalcogenides (TMDs), have attracted intense attention due to their promising properties[1-3]. At the monolayer limit, electrons and holes are confined in these 2D materials, which give rise to a variety of exotic physics phenomena. Monolayer $MoS_2$, with a direct bandgap of 1.8 eV, is an interesting semiconductor material with potential applications in spintronic devices, catalysis and optoelectronic devices[1, 4-6]. To date, chemical vapor deposition (CVD) has shown to be a very effective and feasible approach to obtain monolayer TMDs. It is considered to be the most promising approach for device applications, as large-scale $MoS_2$ growth of high quality, monolayers on $SiO_2/Si$ and other insulating substrates can be achieved[7-9]. However, thickness, crystallinity, defects, and strain depend on the precise control of CVD growth conditions like the temperature, precursor delivery, substrate, pressure, and flow rate. Thus, a straight forward and quick investigation of the properties of CVD grown $MoS_2$ is crucial for understanding the growth and producing tailor-made electronic materials.

The edges of 2D materials are one of their most interesting aspects as they influence the physical properties, including bandgap, superconductivity, mobility, and magnetism[10-14]. It is known that various types of edges have different selectivity towards functionalization with chemical and biological groups[15]. Moreover, catalytic activity has also been found to be related to the edge sites in $MoS_2$[5, 16]. The edge structure is also pivotal for the formation of continuous films in CVD growth of monolayers. However, only STM and HRTEM studies, which are costly, time consuming and require transfer onto suitable substrates, are capable of resolving the edge structure[17]. Only a few studies on the optical properties of the edges in CVD grown $MoS_2$ monolayers exist[18]. Monitoring the peak position, full width at half maximum (FWHM), and intensity of photoluminescence (PL) and Raman have been proven to be fast, non-destructive and effective methods to probe the number of layers, doping, defects, and strain of 2D materials[19-23]. Through PL imaging, variations in

PL intensities have been observed in monolayer TMDs. For the triangular CVD grown monolayer of WS$_2$, very high PL intensities near the edge have been related to structural imperfection and doping[24]. So far, Liu et al. observed the heterogeneity in PL intensities in MoS$_2$ atomic layers and ascribed the changes to strain effects[25]. Further Bao et al. suggested that various PL widths and intensities in MoS$_2$ may be related to S-deficiencies[26].

In this paper, we report spatially resolved PL and Raman studies on CVD grown MoS$_2$ flakes with various shapes. The PL signal (peak position and intensity) and the $E^1_{2g}$ Raman mode at the perimeter are strongly increased in comparison to the center of the flakes, while the FWHM of the A exciton and the position of the $A_{1g}$ Raman mode are uniform over the whole flake. Significantly, the width of the region with increased PL varies between flakes with different edges. Comparisons of as-grown and transferred MoS$_2$ monolayers reveal that non-uniform tensile strain is induced during growth giving rise to the PL and Raman non-uniformities. Our results help to elucidate the fundamental mechanisms contributing to the differences in PL and Raman signals between the perimeter region and the center region of monolayer MoS$_2$, and effects of the induced strain during CVD growth of TMDs.

## II. EXPERIMENTAL SECTION

### A. CVD growth of monolayer MoS$_2$ flakes

The MoS$_2$ flakes were synthesized on 290 nm SiO$_2$/Si in a two-zone CVD furnace using a microcavity reactor as Mo precursor supply. The growth temperature is 750°C with a pressure of 0.9 mbar, as described in detail in Ref[27]. A PMMA support technique was used to transfer the MoS$_2$ flakes onto another SiO$_2$/Si substrate, and details of the method is described as reported[28].

### B. SEM and AFM Characterization

The morphologies and microstructures of the MoS$_2$ flakes were characterized by using scanning electron microscope (SEM, JEOL JSM-6700F) at 10 keV. Atomic force microscopy (AFM, Anasys NanoIR2s) studies were taken in air at ambient condition with tapping mode.

### C. Raman and PL Analysis

Raman and PL spectra were collected using a Witec alpha 300R with a 532 nm excitation laser with a 100 x objective lens (NA=0.95) was at room temperature. The laser power was set at 200 μW for PL measurement in order to avoid laser heating and any other damages produced by the laser. For Raman measurements, a spectral grating with 1800 lines/mm was used whereas for PL the spectral grating was 600 lines/mm. Maps were generated by taking approx. 7 spectra/μm in both x and y directions.

## III. RESULTS AND DISCUSSION

The $MoS_2$ flakes studied in this paper were grown on 290 nm $SiO_2$/Si substrates with a micro-cavity CVD set-up previously described (also see the experimental section) [27]. Fig. 1a provides a typical AFM image of one of the as-grown $MoS_2$ nanoflakes. The identical color contrast manifests uniform thickness of the as-grown $MoS_2$ structure. The thickness of the $MoS_2$ flake was further confirmed by the height profile in the inset of Fig. 1a, it shows that the $MoS_2$ flake has a measured thickness of 0.85 nm, at both edges and varies flat in between, which is in conformity with that of monolayer $MoS_2$ reported by previous studies[29, 30].

For intrinsic monolayer $MoS_2$, two pronounced PL peaks at approximately 1.85 eV (A) and 2.05 eV (B) can be observed. The A peak arises from the neutral exciton emission of the direct transition at the K-point, and the B peak is due to the exciton emission from a second direct transition between the conduction band and a lower-lying valence band[4, 31]. However, numerous factors may contribute to the relative energies and populations of the A exciton in $MoS_2$, including strain, thickness, structural defect densities, and doping[4, 20, 21, 32]. Therefore, detailed PL characterization of the $MoS_2$ flakes has been carried out. A direct comparison of the typical spectra taken at the center and the edge of a flake is shown in Fig. 1b. The charged A exciton at the center shows a peak position of 1.83 eV, and the charged B exciton is at 1.97 eV. The edge site shows a blue shift of about 30 meV to 1.86 eV and 2.01 eV for the A and B exciton, respectively. The map of the peak position of the A exciton in Fig. 1c exhibits a significant contrast between the center region and the perimeter region. This contrast shows an increase of A peak energy from center to perimeter of the monolayer flake. The map of integrated PL intensity of the A peak in Fig. 1d, also shows non-uniformities, with higher intensities at the perimeter. In contrast, decreasing intensity in the outer region was reported by Van der Zande et al.[33]. The inset of Fig. 1c and d present the corresponding line scan of PL position and intensity, respectively, which further confirms the heterogeneity of the PL within a flake.

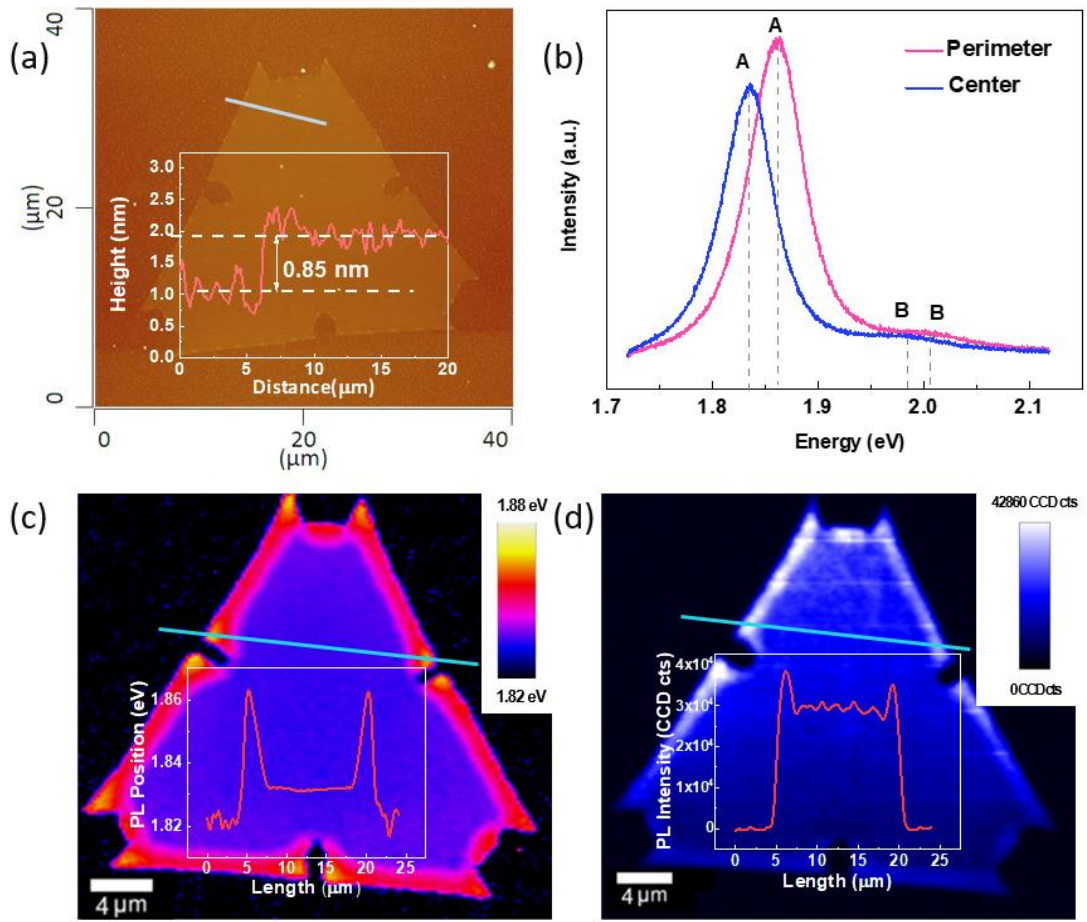

**Fig. 1** (a) AFM image of MoS$_2$ monolayer on SiO$_2$/Si substrate and corresponding height profile (inset) along the blue line. (b) PL spectra for the perimeter region and the center region, respectively. (c) Map of the position of the A exciton of an individual 1L-MoS$_2$ flake and corresponding line scanning profile (inset) along the blue line; (d) Map of the PL intensity of the A exciton of the same flake and corresponding profile (inset) along the blue line.

Raman spectroscopy is a very powerful tool to measure not only the layer number but also the doping and strain in 2D materials[34, 35]. Typically, 2H-MoS$_2$ has two main vibrational Raman modes: E$^1_{2g}$ and A$_{1g}$, where E$^1_{2g}$ represents the in-plane vibration mode of molybdenum and sulfur atoms and A$_{1g}$ is the out-of-plane vibration mode of sulfur atoms[36]. According to previous studies, in typical 2H-type TMDs, the position of A$_{1g}$ mode is known to vary with external electrostatic doping, while the E$^1_{2g}$ mode is sensitive to the strain[34, 35]. Fig. 2a shows a Raman spectrum of E$^1_{2g}$ and A$_{1g}$ modes in the perimeter and in the center of monolayer MoS$_2$, respectively. The two characteristic E$^1_{2g}$ and A$_{1g}$ modes appear at 385.2 cm$^{-1}$ and 404.6 cm$^{-1}$ in the perimeter, while Raman shifts of 383.9 cm$^{-1}$ and 404.5 cm$^{-1}$ are observed in the center. The frequency difference of ~19.4 and 20.6 cm$^{-1}$ between those two peaks agrees well with that of monolayer MoS$_2$. It is noteworthy that the E$^1_{2g}$ mode exhibits a red shift from center to the perimeter, while the A$_{1g}$ peak experiences hardly any shift. To quantitatively analyze the Raman vibration modes, the peak position of the E$^1_{2g}$ and A$_{1g}$ modes are plotted as a function of intensity for several points on the map in the outer and interior region. As we can see in Fig. 2b, it is evident that the E$^1_{2g}$ mode appears in two distinct spectral regions (compared pink and blue data points): 384.5 - 385 cm$^{-1}$ for the perimeter sites and 383.5 - 384 cm$^{-1}$ for the center sites. The relative intensities of the peaks observed at the center and perimeter region are similar. On the other hand, the A$_{1g}$ mode remains at around 404.5 cm$^{-1}$ in all sites. Additionally, we mapped the Raman modes of as-prepared MoS$_2$. The E$^1_{2g}$ peak position across the whole flake is shown in Fig. 2c. This mode varies between the perimeter and the center of the flake. However, as shown in Fig. 2d, the A$_{1g}$ peak position shows spatial uniformities. Since the position of the A$_{1g}$ mode is known to vary with external electrostatic doping, and position of the E$^1_{2g}$ mode is related to the strain, it is proposed that there is no external electrostatic doping, but a significant strain exists across the entire MoS$_2$ layer.

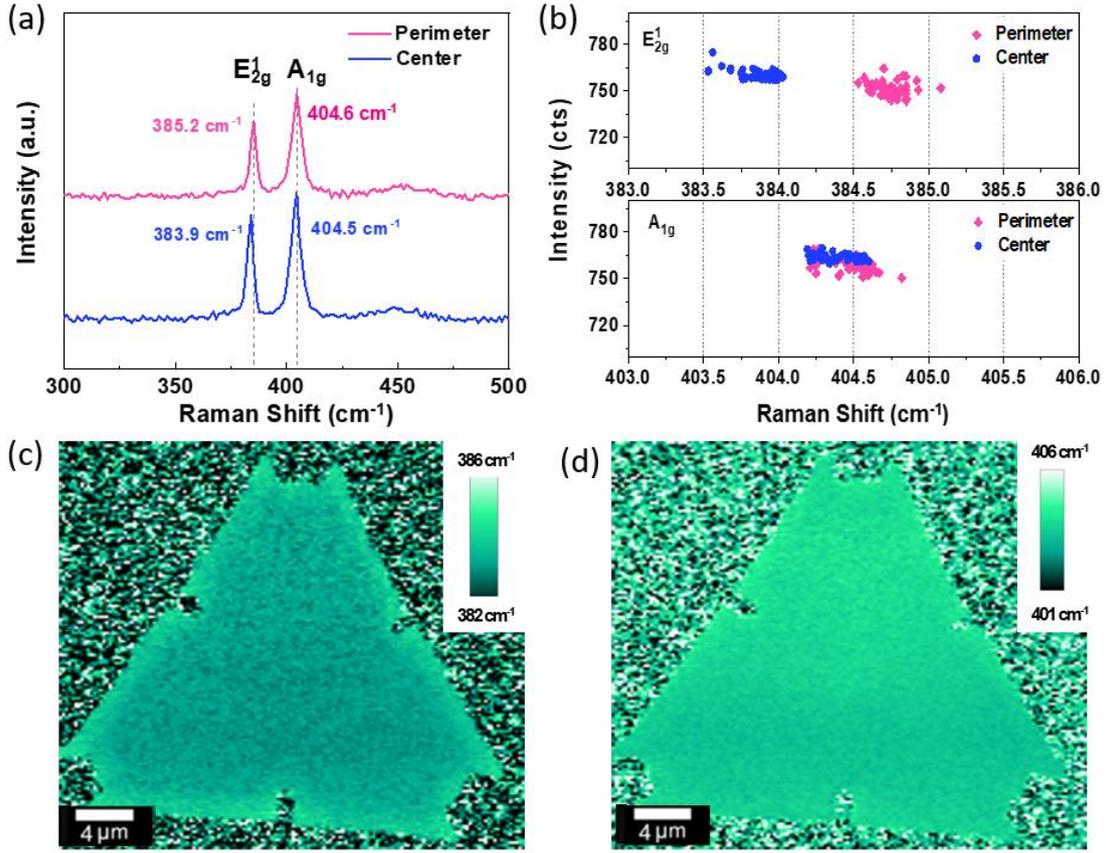

**Fig. 2** (a) Raman spectrum of the perimeter and center region of a $MoS_2$ flake. (b) Raman peak intensities plotted against the peak position in the center (blue) and in the perimeter (pink) region. (c,d) Map of the $E^1_{2g}$ and $A_{1g}$ peak intensity of an individual 1L-$MoS_2$ flake.

According to previous reports, thermally-induced local strain occurs during the growth and may arise from the layer-substrate lattice mismatch. At the end of the growth process, the samples cool rapidly, and the thermal expansion coefficient difference between $SiO_2$ substrate and $MoS_2$ has a significant contraction mismatch. If the as grown $MoS_2$ is much thinner than the substrate, the substrate is essentially stress-free and all the strain will be borne by the grown $MoS_2$[37]. For temperature ranging from 20 °C to 850 °C, the thermal expansion coefficient of $MoS_2$ lattice can be given by [25, 38]:

$$a = 3.1621 + 0.6007 \times 10^{-5}t + 0.3479 \times 10^{-7}t^2 \qquad (1)$$

where $t$ is the sample temperature in ℃. Compared to $MoS_2$, amorphous $SiO_2$ has very low thermal expansion coefficient of ~0.56 ×10-7 /°C², which is too low to affect the calculation result. Thus, by ignoring the value of $SiO_2$, the tensile strain in monolayer

MoS₂ then amounts to:

$$\epsilon = \frac{a_{MoS2}(t=750°C) - a_{MoS2}(t=20°C)}{a_{MoS2}(t=20°C)} \approx 0.76\% \tag{2}$$

On the basis of the growth mechanisms of 2D materials, the growth of MoS₂ will begin from the nucleation center to the perimeter[39]. As a result, the strain field propagates from the center to the circumference, thus, the thermally-induced strain might have a lesser effect on the perimeter of the flakes. Previous works show reduction in A exciton peak energy at a rate of 59 meV/% strain (0.76% strain to ~35 meV reduction), and $E^1_{2g}$ peak shifts at 1.0 ± 1 cm⁻¹%, which aligns quite well with our PL and Raman measurements[40].

To further confirm the thermally-induced strain of the MoS₂ monolayers, we transferred one of the samples onto another SiO₂/Si substrate with a polymer supported wet transfer as described in the experimental section. During this process, the strain of the MoS₂ monolayer can be released. Fig. 3 (a) – (c) are the PL position maps from the A exciton of three MoS₂ flakes with different shapes, after transfer. It can be seen clearly that the transferred MoS₂ flakes present a uniform brightness compared with the flakes before transfer (Fig. 1c), this indicates the strain release after transfer for all the flakes. This supports our observation that the non-uniform PL is originally from the substrate-induce strain during the growth.

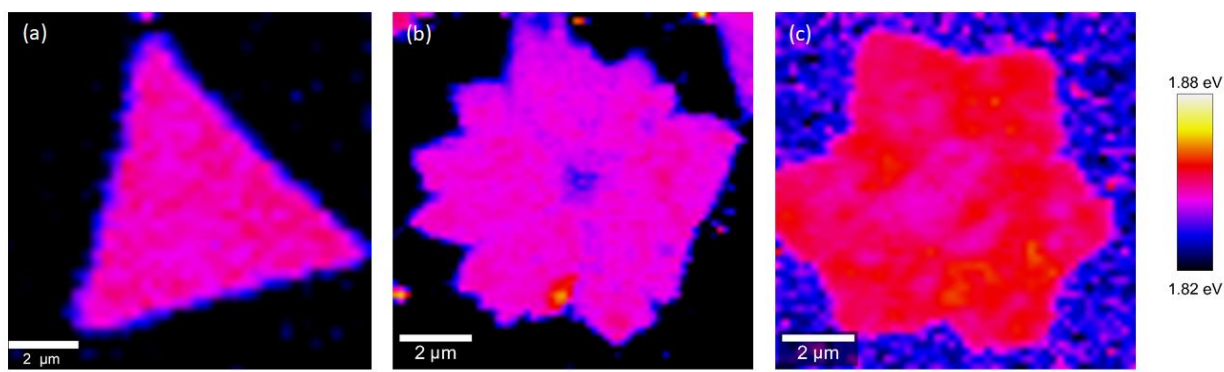

**Fig. 3** Maps of the position of the A exciton of MoS₂ flakes with different shapes, after transfer.

Various shapes of 2D MoS₂ flakes are formed during CVD growth. Fig. 4a and b shows SEM images of monolayer MoS₂ islands in different locations on the same substrate. Fig. 4a shows flakes with straight edges, while Fig. 4b shows others have smooth and less-straight edges, we ascribe the flakes in Fig. 4a as group I (straight edges) and Fig. 4b as group II (smooth edges). The nature of these edges are pivotal for the formation of continuous films in CVD growth and contribute to the properties of the individual flakes. As discussed earlier, outside of STM and HRTEM, identification of the edge type of individual flakes is difficult. While using optical properties of the edges for identification is in its infancy, as a non-destructive method, high resolution PL mapping might be an alternative to distinguish the edge difference since it has been proven to have the ability to identify the grain boundaries (GBs) in the previous studies [17, 33].

It is known that armchair edges have alternating S and Mo atoms, while zigzag edges are terminated by S or Mo only. Van der Zande et al. suggested that from the optical images of $MoS_2$, the sharper, straighter edges are Mo-zigzag terminated[33].

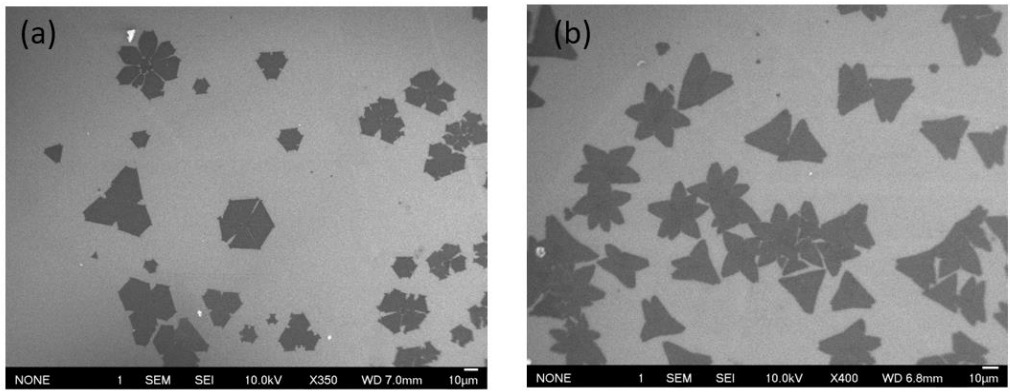

**Fig. 4** Typical SEM images of CVD grown $MoS_2$ sample with (a) sharp and straight edges and (b) smooth and obtuse edges.

Consequently, PL peak position maps were recorded for various flakes in group I and II. The A exciton PL mapping in Fig. 5 a-d (5a and b are from group I, while 5c and d are from group II) shows significant differences between the groups. The width of the area with shifted PL varies for the flakes in group I and II. For every size of flakes, the widths of the shifted-PL perimeter region of group II are wider than those in group I, we will refer to the width of this distinct PL perimeter region as the "perimeter width".

The perimeter widths are estimated to be 1.3 μm for group I, while in group II they are significantly wider at approximately 2.85 μm. In conclusion, high-resolution confocal PL maps not only precisely quantify the difference of the A exciton peak position between the perimeter and the center of MoS$_2$ flakes, but also clearly present that different edge types will give rise to different perimeter widths. In other words, combined with strain effects, the perimeter width of the flakes is indicative of the sample edge type with respect to PL position maps.

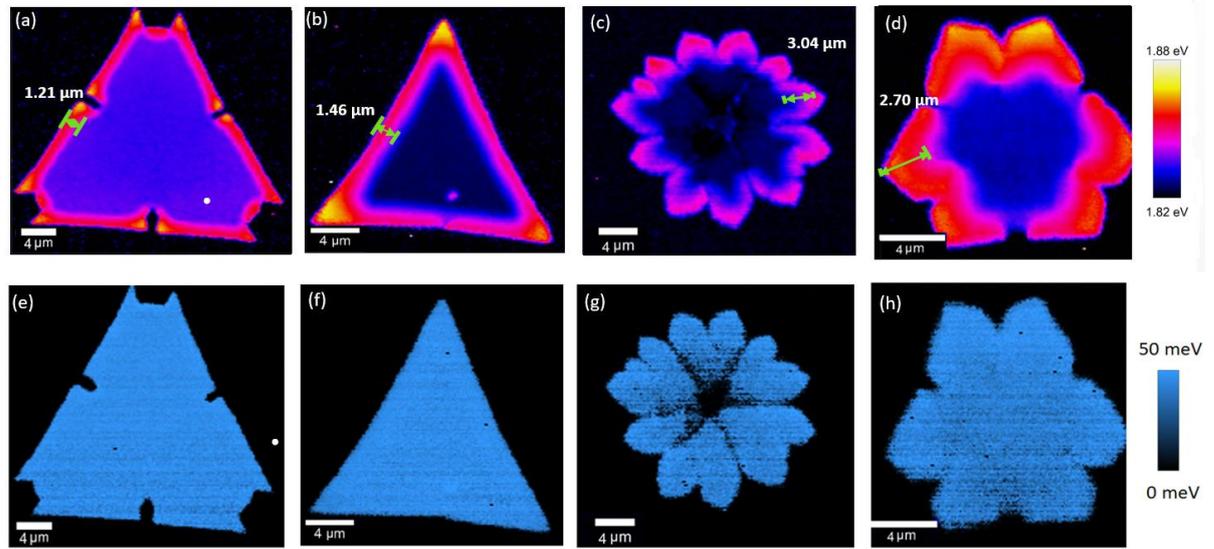

**Fig. 5** (a, b) PL A exciton peak position maps of MoS$_2$ flakes with sharp and straight edges and (c, d) smooth and obtuse edges. (e-h) FWHM maps of the A exciton of MoS$_2$ flakes

Edge defects are also considered to contribute to the full width at half maxima (FWHM) changes of PL[24], however, our PL maps show minimal FWHM changes (Fig. 5e-h). Thus, we exclude edge defects as the main mechanism for the differences observed in the PL responses. That is, thermally-induced strain becomes the likely reason for the variations of the PL in our MoS$_2$ nanoflakes. During the growth process, the strain initiates from the interior and then propagates to the edge. As the growth rate of the armchair, Mo-terminated zigzag, and S-terminated zigzag directions are different[41], the degree of strain relaxation on different edges will also be different. Thus, the different widths of the perimeter region with shifted PL were observed within the two groups. In other words, the changes in PL at the perimeter of the flakes could not arise from the edge state or defects alone, the thermal expansion coefficient mismatch of the substrate and the induced strain in the center of the flake alongside different strain relaxation rates of each edge type makes the most significant contribution to the observed PL maps

## IV. CONCLUSION

In this report, MoS$_2$ monolayer flakes with various shapes grown by CVD were characterized by PL and Raman maps to study the structural heterogeneity and strain effects. The PL signal peak position, intensity, and the $E^1_{2g}$ Raman mode position show significant differences at the edges and in the interior of monolayer flakes, while the FWHM of the A exciton and the $A_{1g}$ Raman mode are uniform. This indicates an intrinsic tensile strain in MoS$_2$ which likely is induced during the fast cooling process after growth.

For different edge types (straight or smooth edge), different widths of the perimeter region have been observed. This difference may be due to various strain relaxation rates in MoS$_2$ along the growth direction. Our results suggest that thermally induced non-uniform tensile strain plays a significant role in CVD growth of MoS$_2$ monolayers, and also helps to elucidate the fundamental mechanisms responsible for differences between the perimeter and center of PL and Raman in monolayer MoS$_2$.

**Acknowledgments**

G. S. D. acknowledges the support of SFI under Contract No. PI_15/IA/3131 and the European Commission under Grant Agreement n°785219. This work was also supported by National Natural Science Foundation of China (No. 11474244), the National Basic Research Program of China (2015CB921103) Science, Technology Program of Xiangtan (No. CXY-ZD20172002), and Innovative Research Team in University (IRT13093).

**References**

[1] K.F. Mak, C. Lee, J. Hone, J. Shan, T.F. Heinz, Atomically thin MoS 2: a new direct-gap semiconductor, Phys. Rev.Lett., 105 (2010) 136805.

[2] G.-C. Guo, D. Wang, X.-L. Wei, Q. Zhang, H. Liu, W.-M. Lau, L.-M. Liu, Principles study of phosphorene and graphene heterostructure as anode materials for rechargeable Li batteries, J. Phys. Chem. Lett., 6 (2015) 5002-5008.

[3] K.S. Novoselov, A.K. Geim, S. Morozov, D. Jiang, M. Katsnelson, I. Grigorieva, S. Dubonos, Firsov, AA, Two-dimensional gas of massless Dirac fermions in graphene, Nature, 438 (2005) 197.

[4] A. Splendiani, L. Sun, Y. Zhang, T. Li, J. Kim, C.-Y. Chim, G. Galli, F. Wang, Emerging photoluminescence in monolayer $MoS_2$, Nano Lett., 10 (2010) 1271-1275.

[5] H. Nolan, N. McEvoy, M. O'Brien, N.C. Berner, C. Yim, T. Hallam, A.R. McDonald, G.S. Duesberg, Molybdenum disulfide/pyrolytic carbon hybrid electrodes for scalable hydrogen evolution, Nanoscale, 6 (2014) 8185-8191.

[6] J. Pandey, A. Soni, Unraveling biexciton and excitonic excited states from defect bound states in monolayer $MoS_2$, Appl. Surf. Sci., 463 (2019) 52-57.

[7] W. Zhang, J.K. Huang, C.H. Chen, Y.H. Chang, Y.J. Cheng, L.J. Li, High-gain phototransistors based on a CVD $MoS_2$ monolayer, Adv. Mater., 25 (2013) 3456-3461.

[8] H. Li, Y. Li, A. Aljarb, Y. Shi, L.-J. Li, Epitaxial growth of two-dimensional layered transition-metal dichalcogenides: growth mechanism, controllability, and scalability, Chem. Rev., 118 (2017) 6134-6150.

[9] J. Shan, J. Li, X. Chu, M. Xu, F. Jin, X. Fang, Z. Wei, X. Wang, Enhanced photoresponse characteristics of transistors using CVD-grown $MoS_2/WS_2$ heterostructures, Appl. Surf. Sci., 443 (2018) 31-38.

[10] C. Zhang, A. Johnson, C.-L. Hsu, L.-J. Li, C.-K. Shih, Direct imaging of band profile in single layer $MoS_2$ on graphite: quasiparticle energy gap, metallic edge states, and edge band bending, Nano Lett., 14 (2014) 2443-2447.

[11] N.F. Yuan, K.F. Mak, K. Law, Possible topological superconducting phases of MoS 2, Phys. Rev.Lett., 113 (2014) 097001.

[12] Y. Cai, G. Zhang, Y.-W. Zhang, Polarity-reversed robust carrier mobility in

monolayer MoS$_2$ nanoribbons, J. Am. Chem. Soc., 136 (2014) 6269-6275.

[13] J. Zheng, X. Yan, Z. Lu, H. Qiu, G. Xu, X. Zhou, P. Wang, X. Pan, K. Liu, L. Jiao, High-mobility multilayered MoS$_2$ flakes with low contact resistance grown by chemical vapor deposition, Adv. Mater., 29 (2017) 1604540.

[14] A. Vojvodic, B. Hinnemann, J.K. Nørskov, Magnetic edge states in MoS$_2$ characterized using density-functional theory, Phys. Rev. B, 80 (2009) 125416.

[15] Z. Yin, H. Li, H. Li, L. Jiang, Y. Shi, Y. Sun, G. Lu, Q. Zhang, X. Chen, H. Zhang, Single-layer MoS$_2$ phototransistors, ACS Nano, 6 (2011) 74-80.

[16] T.F. Jaramillo, K.P. Jørgensen, J. Bonde, J.H. Nielsen, S. Horch, I. Chorkendorff, Identification of active edge sites for electrochemical H$_2$ evolution from MoS$_2$ nanocatalysts, Science, 317 (2007) 100-102.

[17] K. Santosh, R.C. Longo, R. Addou, R.M. Wallace, K. Cho, Impact of intrinsic atomic defects on the electronic structure of MoS$_2$ monolayers, Nanotechnology, 25 (2014) 375703.

[18] T.H. Ly, S.J. Yun, Q.H. Thi, J. Zhao, Edge delamination of monolayer transition metal dichalcogenides, ACS Nano, 11 (2017) 7534-7541.

[19] H. Li, Q. Zhang, C.C.R. Yap, B.K. Tay, T.H.T. Edwin, A. Olivier, D. Baillargeat, From bulk to monolayer MoS$_2$: evolution of Raman scattering, Adv. Funct. Mater., 22 (2012) 1385-1390.

[20] S. Mouri, Y. Miyauchi, K. Matsuda, Tunable photoluminescence of monolayer MoS$_2$ via chemical doping, Nano letters, 13 (2013) 5944-5948.

[21] H. Nan, Z. Wang, W. Wang, Z. Liang, Y. Lu, Q. Chen, D. He, P. Tan, F. Miao, X. Wang, Strong photoluminescence enhancement of MoS2 through defect engineering and oxygen bonding, ACS Nano, 8 (2014) 5738-5745.

[22] A. Castellanos-Gomez, R. Roldán, E. Cappelluti, M. Buscema, F. Guinea, H.S. van der Zant, G.A. Steele, Local strain engineering in atomically thin MoS$_2$, Nano Lett., 13 (2013) 5361-5366.

[23] S. Kataria, S. Wagner, T. Cusati, A. Fortunelli, M.C. Lemme, Growth-induced strain in chemical vapor deposited monolayer MoS$_2$: experimental and theoretical investigation, Adv. Mater. Interfaces, 4 (2017) 1700031.

[24] H.R. Gutiérrez, N. Perea-López, A.L. Elías, A. Berkdemir, B. Wang, R. Lv, F. López-Urías, V.H. Crespi, H. Terrones, M. Terrones, Extraordinary room-temperature photoluminescence in triangular $WS_2$ monolayers, Nano Lett., 13 (2012) 3447-3454.

[25] Z. Liu, M. Amani, S. Najmaei, Q. Xu, X. Zou, W. Zhou, T. Yu, C. Qiu, A.G. Birdwell, F.J. Crowne, Strain and structure heterogeneity in $MoS_2$ atomic layers grown by chemical vapour deposition, Nat. Commun., 5 (2014) 5246.

[26] W. Bao, N.J. Borys, C. Ko, J. Suh, W. Fan, A. Thron, Y. Zhang, A. Buyanin, J. Zhang, S. Cabrini, Visualizing nanoscale excitonic relaxation properties of disordered edges and grain boundaries in monolayer molybdenum disulfide, Nat. Commun., 6 (2015) 7993.

[27] M. O'Brien, N. McEvoy, T. Hallam, H.-Y. Kim, N.C. Berner, D. Hanlon, K. Lee, J.N. Coleman, G.S. Duesberg, Transition metal dichalcogenide growth via close proximity precursor supply, Sci. Rep., 4 (2014) 7374.

[28] M. O'Brien, N. Scheuschner, J. Maultzsch, G.S. Duesberg, N. McEvoy, Raman Spectroscopy of Suspended $MoS_2$, Phys. Status Solidi (b), 254 (2017) 1700218.

[29] G. Tai, T. Zeng, J. Yu, J. Zhou, Y. You, X. Wang, H. Wu, X. Sun, T. Hu, W. Guo, Fast and large-area growth of uniform $MoS_2$ monolayers on molybdenum foils, Nanoscale, 8 (2016) 2234-2241.

[30] W. Zhu, T. Low, Y.-H. Lee, H. Wang, D.B. Farmer, J. Kong, F. Xia, P. Avouris, Electronic transport and device prospects of monolayer molybdenum disulphide grown by chemical vapour deposition, Nat. Commun., 5 (2014) 3087.

[31] S. Wang, X. Wang, J.H. Warner, All chemical vapor deposition growth of $MoS_2$: h-BN vertical van der Waals heterostructures, ACS Nano, 9 (2015) 5246-5254.

[32] K. He, C. Poole, K.F. Mak, J. Shan, Experimental demonstration of continuous electronic structure tuning via strain in atomically thin $MoS_2$, Nano Lett., 13 (2013) 2931-2936.

[33] A.M. Van Der Zande, P.Y. Huang, D.A. Chenet, T.C. Berkelbach, Y. You, G.-H. Lee, T.F. Heinz, D.R. Reichman, D.A. Muller, J.C. Hone, Grains and grain boundaries in highly crystalline monolayer molybdenum disulphide, Nat. Mater., 12 (2013) 554.

[34] M. Rahaman, R.D. Rodriguez, G. Plechinger, S. Moras, C. Schüller, T. Korn, D.R.

Zahn, Highly localized strain in a MoS$_2$/Au Heterostructure revealed by tip-enhanced Raman spectroscopy, Nano Lett., 17 (2017) 6027-6033.

[35] B. Chakraborty, A. Bera, D. Muthu, S. Bhowmick, U.V. Waghmare, A. Sood, Symmetry-dependent phonon renormalization in monolayer MoS$_2$ transistor, Phys. Rev. B, 85 (2012) 161403.

[36] T. Wieting, J. Verble, Infrared and Raman studies of Long-Wavelength optical phonons in hexagonal MoS$_2$, Phys. Rev. B, 3 (1971) 4286.

[37] V. Srikant, D. Clarke, Optical absorption edge of ZnO thin films: the effect of substrate, J.Appl. Phys., 81 (1997) 6357-6364.

[38] S. El-Mahalawy, B. Evans, The thermal expansion of 2H-MoS$_2$, 2H-MoSe$_2$ and 2H-WSe$_2$ between 20 and 800°C, J. Appl. Crystallogr, 9 (1976) 403-406.

[39] S. Wang, Y. Rong, Y. Fan, M. Pacios, H. Bhaskaran, K. He, J.H. Warner, Shape evolution of monolayer MoS2 crystals grown by chemical vapor deposition, Chem. of Mater., 26 (2014) 6371-6379.

[40] H.J. Conley, B. Wang, J.I. Ziegler, R.F. Haglund Jr, S.T. Pantelides, K.I. Bolotin, Bandgap engineering of strained monolayer and bilayer MoS$_2$, Nano Lett., 13 (2013) 3626-3630.

[41] S. Hao, B. Yang, Y. Gao, Unravelling merging behaviors and electrostatic properties of CVD-grown monolayer MoS$_2$ domains, J. Chem. Phys., 145 (2016) 084704.